\documentclass[a4paper,12pt]{article}
\usepackage{amsmath}
\usepackage{amsfonts}
\usepackage{amssymb}

\title{Effects of Ill-Defined Domain of Definitions of the Parameter Operator on Berry Curvature and the Adiabatic Theorem}
\author{Georgios Konstantinou, Konstantinos Moulopoulos}
\date{\today}
\begin{document}

\maketitle

\begin{abstract}
We present a comprehensive analytical study that extends the conventional formulation of Berry curvature, highlighting its derivation in the context of problematic domains of definition of the operators. Our analysis reveals that handling these domains carefully can have a substantial impact on Berry curvature, demonstrating that even Hamiltonians without explicit parameter dependence may exhibit nonzero Berry curvature. This finding emphasizes that Berry curvature is intrinsically related to the eigenvectors rather than the Hamiltonian itself. Our approach utilizes the standard Bloch (\( \vec{k} \)-space) framework for spatially periodic systems, illustrating these effects from first principles and discussing potential implications for solid-state systems.
\end{abstract}

\section{Introduction}
In time-dependent quantum systems, the adiabatic theorem asserts that adiabaticity is preserved if no inter-level transitions occur. Specifically, if a system is initially in a single eigenstate, it will remain in that eigenstate throughout the parameter evolution until the final time \( T \) (which in the literature corresponds to a cyclic process). Each eigenstate \( | n(\vec{R}) \rangle \), assumed single-valued in parameter space, is defined by the eigenvalue equation:
\begin{equation}
H(\vec{R}) | n(\vec{R}) \rangle = \epsilon_n(\vec{R}) | n(\vec{R}) \rangle
\end{equation}
where \( \vec{R} = \vec{R}(t) \) is a collective time-dependent parameter vector, valid for all times \( t \in [0, T] \). The adiabaticity condition (no transitions to other states \( n+1, n-1, \ldots \)) holds due to the vanishing of the rightmost term in the following equation, derived from substituting \( |\Psi \rangle = \sum_{n} c_n(t) | n \rangle \) into the Schrödinger equation:
\begin{equation}
\dot{c}_m(t) = c_m(t) \left(-i \frac{\epsilon_m}{\hbar} - \langle m | \dot{m} \rangle \right) - i \hbar \sum_{n \ne m} c_n(t) \langle m | \dot{n} \rangle
\end{equation}
Here, \( \langle m | \dot{n} \rangle \ll 1 \) is the standard condition for adiabaticity. Several studies have attempted to quantify 'how small', with results often being contradictory \cite{Tong2005}. By manipulating this term, one incorporates the time derivative of the Hamiltonian into the adiabaticity condition:
\begin{equation}
\dot{H} | n \rangle + H | \dot{n} \rangle = \dot{\epsilon}_n | n \rangle + \epsilon_n | \dot{n} \rangle
\end{equation}
Taking the inner product with \( \langle m | \), with \( m \ne n \), and using orthonormality, we find:
\begin{equation}
\langle m | \dot{H} | n \rangle + \langle m | H | \dot{n} \rangle = \epsilon_n \langle m | \dot{n} \rangle
\end{equation}
For a Hermitian Hamiltonian \( H = H^\dagger \), we have:
\begin{equation}
\langle m | H | \dot{n} \rangle = \epsilon_m \langle m | \dot{n} \rangle
\end{equation}
so the adiabatic transition matrix element becomes:
\begin{equation}
\langle m | \dot{n} \rangle = \frac{\langle m | \dot{H} | n \rangle}{\epsilon_n - \epsilon_m}
\end{equation}
When time dependence arises only through parameters, the derivative becomes:
\begin{equation}
\langle m | \nabla_{\vec{R}} | n \rangle = \frac{\langle m | \nabla_{\vec{R}} H | n \rangle}{\epsilon_n - \epsilon_m}
\end{equation}
The Berry connection and curvature are then defined as \cite{berry1984}:
\begin{equation}
\vec{A} = i \langle n | \nabla_{\vec{R}} | n \rangle
\end{equation}
\begin{equation}
\vec{\Omega} = \nabla_{\vec{R}} \times \vec{A}
\end{equation}
An alternative expression for Berry curvature is:
\begin{equation}
\vec{\Omega} = i \sum_{m \ne n} \langle \nabla_{\vec{R}} n | m \rangle \times \langle m | \nabla_{\vec{R}} n \rangle
\end{equation}
Using Eq. (7), we obtain:
\begin{equation}
\vec{\Omega} = i \sum_{m \ne n} \frac{\langle n | \nabla_{\vec{R}} H | m \rangle \times \langle m | \nabla_{\vec{R}} H | n \rangle}{(\epsilon_n - \epsilon_m)^2}
\end{equation}
This standard form suggests that Berry curvature depends on the Hamiltonian and its derivatives. However, in cases where \( H \) is parameter-independent (as in the standard Bloch Hamiltonian in solids), the right-hand side of Eq. (11) is zero. This contradicts the fact that Berry curvature is often non-zero in such systems.

To resolve this, we revisit the derivation and note subtleties arising from the domain of definition of the operators involved. Specifically, the conventional assumption that all operators act on a common, well-defined domain can break down in parameter-dependent systems. In particular, while the Hamiltonian \( H \) remains Hermitian throughout, the parameter-derivative operator \( \nabla_{\vec{R}} \) may act on eigenstates \( |n(\vec{R})\rangle \) that lie outside the domain of \( H \). This mismatch leads to nontrivial corrections when evaluating matrix elements involving \( \nabla_{\vec{R}} \).

To account for these domain-related subtleties, we define a correction term:
\begin{equation}
\vec{\Delta}_{m,n} = \langle m | H | \nabla_{\vec{R}} n \rangle - \langle H m | \nabla_{\vec{R}} n \rangle
\end{equation}
This term vanishes only when the action of \( H \) on \( | m \rangle \) and its adjoint action on \( | \nabla_{\vec{R}} n \rangle \) are both well-defined and compatible—i.e., when the relevant wavefunctions and their parameter derivatives all reside within the domain of \( H \). Therefore, \(\vec{\Delta}_{m,n}\) does not signal any failure of Hermiticity of the Hamiltonian itself, but rather exposes an inconsistency in the domain of the parameter derivative operator.

A physically important and illustrative example is when the parameter \( \vec{R} \) is the crystal momentum \( \vec{k} \) in a periodic solid. The standard Bloch Hamiltonian \( H(\vec{k}) \), often written in the form:
\[
H(\vec{k}) = \frac{1}{2m} \left( -i\hbar \nabla + \hbar \vec{k} \right)^2 + V(\vec{r})
\]
is commonly treated as Hermitian for each \( \vec{k} \). However, the operation \( \nabla_{\vec{k}} | u_{n,\vec{k}} \rangle \), where \( | u_{n,\vec{k}} \rangle \) is the periodic part of the Bloch function, can produce a vector that lies outside the domain of \( H \), particularly due to boundary or gauge ambiguities associated with the Bloch wavefunctions.

In this case, even though \( H(\vec{k}) \) itself has no explicit \( \vec{k} \)-derivative terms and thus \( \nabla_{\vec{k}} H = 0 \), the Berry curvature \( \vec{\Omega}_n(\vec{k}) \) can be nonzero due to nonvanishing surface or boundary-like contributions encapsulated by \( \vec{\Delta}_{m,n} \). These terms reflect how parameter variations affect the structure of the eigenstates, even when the Hamiltonian is independent of the parameter in a formal sense.

This analysis demonstrates that Berry curvature is fundamentally an eigenstate property, and its nontrivial structure can originate from subtle domain issues in the differentiation with respect to parameters—especially in systems with periodicity, boundary constraints, or gauge freedoms.

This can be expressed using wavefunctions \( f_n \) in position representation as:
\begin{equation}
\vec{\Delta}_{m,n} = \frac{\hbar^2}{2m} \int \nabla \left( \nabla f^*_m \nabla_{\vec{R}} f_n - f^*_m \nabla \nabla_{\vec{R}} f_n - 2i \frac{e}{\hbar c} f^*_m \vec{A} \nabla_{\vec{R}} f_n \right) d^3 r
\end{equation}
The quantity in parentheses is a generalized current density:
\begin{equation}
\vec{J}_{m,n} = \frac{\hbar^2}{2m} \left( \nabla f^*_m \nabla_{\vec{R}} f_n - f^*_m \nabla \nabla_{\vec{R}} f_n - 2i \frac{e}{\hbar c} f^*_m \vec{A} \nabla_{\vec{R}} f_n \right)
\end{equation}
Thus, the correction becomes a surface term:
\begin{equation}
\vec{\Delta}_{m,n} = \oint \vec{J}_{m,n} \cdot d\vec{S}
\end{equation}
Incorporating this correction modifies the matrix element:
\begin{equation}
\langle m | \nabla_{\vec{R}} | n \rangle = \frac{\langle m | \nabla_{\vec{R}} H | n \rangle + \vec{\Delta}_{m,n}}{\epsilon_n - \epsilon_m}
\end{equation}
and the Berry curvature becomes:
\begin{equation}
\vec{\Omega}^n = i \sum_{m \ne n} \frac{(\langle n | \nabla_{\vec{R}} H | m \rangle + \vec{\Delta}_{n,m}) \times (\langle m | \nabla_{\vec{R}} H | n \rangle + \vec{\Delta}_{m,n})}{(\epsilon_n - \epsilon_m)^2}
\end{equation}
This expression restores nonzero Berry curvature even in the absence of explicit parameter dependence in \( H \), thus resolving the inconsistency. It confirms that Berry curvature fundamentally arises from the eigenvectors and their boundary behaviors, and not just from explicit parametric dependence in the Hamiltonian.

\section{Revised Adiabatic Theorem}
Referring to Equation (2) for the adiabatic theorem, and exploring the generalized adiabaticity conditions as previously mentioned, we note that the following relation is also valid:
\begin{equation}
\langle m | \dot{n} \rangle = \frac{\langle m | \dot{H} | n \rangle + \Delta_{m,n}}{\epsilon_n - \epsilon_m}
\end{equation}
Thus, the extended version of the ordinary adiabaticity condition must be:
\begin{equation}
\left| \frac{\langle m | \dot{H} | n \rangle + \Delta_{m,n}}{\epsilon_n - \epsilon_m} \right| \ll 1
\end{equation}
This highlights the necessity of incorporating the corrective term \( \Delta_{m,n} \), which originates not from any non-Hermitian behavior of \( H \), but rather from subtleties in the domain of definition of the operator \( \nabla_{\vec{R}} \), or more generally, of the parameter-derivative operator acting on the eigenstates. This is especially critical in systems where \( \dot{H} \) vanishes — such as the Bloch problem — and yet adiabatic transport still occurs.

For illustration, consider the standard Bloch problem with the Hamiltonian:
\begin{equation}
H = \frac{p^2}{2m} + V(\vec{r})
\end{equation}
where \( V(\vec{r}) \) is the periodic potential of the crystal. The eigenfunctions of this Hamiltonian are the well-known Bloch states: \( f_{n,\vec{k}}(\vec{r}) = u_{n,\vec{k}}(\vec{r}) \exp(i \vec{k} \cdot \vec{r}) \), with \( u_{n,\vec{k}} \) being the cell-periodic functions. Importantly, the Hamiltonian \( H \) does not explicitly depend on the crystal momentum \( \vec{k} \); rather, the \( \vec{k} \)-dependence appears entirely in the eigenfunctions.

If one were to compute adiabaticity conditions using only \( \langle m | \dot{H} | n \rangle \), the result would be zero, incorrectly suggesting a lack of dynamical response to changes in \( \vec{k}(t) \). However, using the corrected Equation (18), which includes \( \Delta_{m,n} \), yields a meaningful, non-zero result even in the absence of explicit time-dependence in \( H \). This correction captures the nontrivial role played by the domains of the derivative operator \( \nabla_{\vec{k}} \), which acts on wavefunctions that may lie outside the strict domain of \( H \).

For simplicity, considering a one-dimensional Bloch solid, the following explicit result is obtained (see Appendix A for the derivation):
\begin{equation}
\Delta_{n'k',nk} = \frac{\hbar}{m} \left[\langle p \rangle_{n'k',nk} + \frac{im}{\hbar} (\epsilon_{nk} - \epsilon_{n'k'}) \langle x \rangle_{n'k',nk} \right]
\end{equation}
where \( \langle p \rangle_{n'k',nk} \) and \( \langle x \rangle_{n'k',nk} \) are the off-diagonal elements of the momentum and position operators, respectively.

It is noteworthy that when using the explicitly \( \vec{k} \)-dependent Hamiltonian \( H(k) \) instead, the correction term \( \Delta \) vanishes. This is due to the periodicity of the cell-periodic functions \( u_{n,k} \) and the smooth behavior of their derivatives with respect to \( x \) and \( k \) (see, for instance, Ref. [4] and comments in Appendix C). In this case, one finds:
\begin{equation}
\langle n'k' | \dot{H}(k) | nk \rangle = \frac{\hbar}{m} \langle p \rangle_{n'k',nk}
\end{equation}
This expression, differing from Equation (21), reflects that the two formulations of the Bloch problem — using \( H \) or \( H(k) \) — are not strictly equivalent from the standpoint of operator domains. While the energy spectrum remains invariant, the eigenfunctions and the operator structure do not. In the context of the adiabatic theorem, these differences are crucial: one must account for the subtleties introduced by the parameter-derivative operator acting outside the native domain of the unparameterized Hamiltonian \( H \). This distinction, though subtle, has significant implications for the validity and application of adiabaticity conditions in systems such as periodic solids.

\section{Extended Berry Curvature Considerations: Gauge Independence}
From the above, we observe that the dependence of $\vec{\Omega}^n$ on the eigenvectors $|n(\vec{R})\rangle$ is indeed a crucial element, as it actually was for the initial form of the Berry curvature, as given by  Eq. (9), had we not assumed single-valued bands. This actually reflects the fact that by transitioning from Eq. (9) to Eq. (17), one transfers any abnormalities related to the eigenvectors to Eq. (17), unavoidably, (through the $\vec{\Delta}$ term). Therefore. the non-Hermitian term, in reality, restores the the meaning of the extended formula as of being a Berry curvature. Additionally, the condition $\sum_{n} \vec{\Omega}^n = 0$ remains valid: when one carefully writes down Eq. (17) into its individual components, these terms cancel out symmetrically. For instance, consider the following pairs of terms appearing in the numerator of the $z$-component of $\vec{\Omega}^n$ (Eq. 17):
\begin{equation}
\langle n \left| \frac{\partial H}{\partial R_x} \right| m \rangle \langle m \left| \frac{\partial H}{\partial R_y} \right| n \rangle - \langle n \left| \frac{\partial H}{\partial R_y} \right| m \rangle \langle m \left| \frac{\partial H}{\partial R_x} \right| n \rangle
\end{equation}
\begin{align}
(\Delta_{n,m})_x \langle m \left| \frac{\partial H}{\partial R_y} \right| n \rangle 
- \langle n \left| \frac{\partial H}{\partial R_y} \right| m \rangle (\Delta_{m,n})_x & \nonumber \\
+ \langle n \left| \frac{\partial H}{\partial R_x} \right| m \rangle (\Delta_{m,n})_y 
- (\Delta_{n,m})_y \langle m \left| \frac{\partial H}{\partial R_x} \right| n \rangle & 
\end{align}
\begin{equation}
(\Delta_{n,m})_x (\Delta_{m,n})_y - (\Delta_{n,m})_y (\Delta_{m,n})_x
\end{equation}
It is straightforward to see that summing over $\sum_{n} \Omega^n_z$ results in zero, due to the vanishing of symmetric terms appearing in the above equations (the same result holds for all components of $\vec{\Omega}^n$), leading us to the local conservation law: $\sum_{n} \vec{\Omega}^n = 0$.
Finally, let us address the question of gauge invariance of this curvature. To verify this, consider the transformation: $|n'\rangle \rightarrow e^{i\beta_n(\vec{R})}|n\rangle$ and $|m'\rangle \rightarrow e^{i\beta_m(\vec{R})}|m\rangle$. Under this transformation, $\vec{\Delta}_{m,n}$ transforms as (from Eq. 12, with $m$ being a dummy index in the summation of Eq. (17)):
\begin{equation}
\vec{\Delta}'_{m,n} = \langle m' \left| H \nabla_{\vec{R}} \right| n' \rangle - \langle H m' \left| \nabla_{\vec{R}} \right| n' \rangle
\end{equation}
\begin{equation}
=e^{i(\beta_n - \beta_m)} \left[ i(\epsilon_n - \epsilon_m) \nabla \beta_n \langle m | n \rangle + \langle m | H \nabla n \rangle + \epsilon_m \langle m | \nabla n \rangle \right]
\end{equation}
\begin{equation}
=e^{i(\beta_n - \beta_m)} \left( \langle m | H \nabla n \rangle + \langle Hm | \nabla n \rangle \right) = e^{i(\beta_n - \beta_m)} \Delta_{(m,n)}
\end{equation}
due to the orthogonality of the eigenstates. Similarly, we find:
\begin{equation}
\vec{\Delta}'_{n,m} =e^{i(\beta_m - \beta_n)} \Delta_{(n,m)}
\end{equation}
Using then Eq. (17), we obtain the expected result:
\begin{equation}
\vec{\Omega}'^n = \vec{\Omega}^n
\end{equation}
We thus conclude that any irregularities arising from, e.g., multi-valued phase factors (even though Eq. (17) still involves gradients of the eigenvectors, preserving the initial definition of the Berry curvature), are eliminated by the orthogonality of the eigenstates, resulting in a gauge-invariant Berry curvature.

\section{Application to the Free Particle in 3 Dimensions}
We start by applying Eq. (16) to the simple case of a free particle with Hamiltonian:
\begin{equation}
H = \frac{p^2}{2m}
\end{equation}
In this simplest case, the Hamiltonian does not depend on any parameters (i.e., it is independent of $\vec{k}$). The eigenfunctions are, under specific considerations, plane waves of the form $f_{\vec{k}} = e^{i\vec{k} \cdot \vec{r}}$, and the eigenenergies are given by:
\begin{equation}
\epsilon_{\vec{k}} = \frac{\hbar^2 k^2}{2m}
\end{equation}
By straightforward calculation of the matrix element in Eq. (16) (with $m=k', n=k$ and $\vec{R}=\vec{k}$, we obtain:
\begin{equation}
\langle \vec{k}' | \nabla_{\vec{k}} | \vec{k} \rangle = i \int d^3r \, \vec{r} \, e^{i(\vec{k} - \vec{k}') \cdot \vec{r}}
\end{equation}
For consistency, we may use Eq. (16) to check the effects of non-Hermiticity. Given that $\nabla_{\vec{k}} H = 0$, Eq. (16) becomes:
\begin{equation}
\langle \vec{k}' | \nabla_{\vec{k}} | \vec{k} \rangle = \frac{\vec{\Delta}_{\vec{k}', \vec{k}}}{\epsilon_{\vec{k}} - \epsilon_{\vec{k}'}}
\end{equation}
where the non-Hermitian contribution term, derived from Eq. (13) with $m = \vec{k}'$, $n = \vec{k}$, $\vec{R} = \vec{k}$, and vanishing vector potential, is:
\begin{equation}
\vec{\Delta}_{\vec{k}', \vec{k}} = \frac{\hbar^2}{2m} \int \nabla \left( \nabla f^*_{\vec{k}'} \nabla_{\vec{k}} f_{\vec{k}} - f^*_{\vec{k}'} \nabla \nabla_{\vec{k}} f_{\vec{k}} \right) d^3r
\end{equation}
After carrying out the necessary steps, one can easily arrive at the following result:
\begin{equation}
\vec{\Delta}_{\vec{k}', \vec{k}} = \frac{\hbar^2}{2m} \int d^3r \left( 2 \vec{k} \, e^{i(\vec{k} - \vec{k}') \cdot \vec{r}} + i (k^2 - k'^2) \vec{r} \, e^{i(\vec{k} - \vec{k}') \cdot \vec{r}} \right)
\end{equation}
Here, the first term vanishes due to orthonormality conditions. Thus, Eq. (16) simplifies to:
\begin{equation}
\langle \vec{k}' | \nabla_{\vec{k}} | \vec{k} \rangle = \frac{\frac{\hbar^2}{2m} i (k^2 - k'^2) \int d^3r \, \vec{r} \, e^{i(\vec{k} - \vec{k}') \cdot \vec{r}}}{\frac{\hbar^2}{2m} (k^2 - k'^2)} = i \int d^3r \, \vec{r} \, e^{i(\vec{k} - \vec{k}') \cdot \vec{r}}
\end{equation}
which exactly coincides with the expected result from Eq. (33), showcasing the necessity of including the non-Hermitian terms in the calculations. 

\section{Application to Bloch Solids}

Following the same logic as shown in Eq.(16) regarding the appearance of an additional term in the off-diagonal matrix elements, similar considerations apply to Berry curvature calculations. For clarity, we consider the example of a Bloch solid described by the Hamiltonian:
\begin{equation}
H(\vec{k}) = \frac{p^2}{2m} + V(\vec{r})
\end{equation}
with eigenfunctions \( f_{n, \vec{k}} = u_{n, \vec{k}}(\vec{r}) e^{i \vec{k} \cdot \vec{r}} \).

Once again, we encounter a situation where the Hamiltonian does not explicitly depend on the parameter \(\vec{k}\), such that \( \nabla_{\vec{k}} H = 0 \). In this case, if one applies Eq. (11) to calculate the Berry curvature, the result would incorrectly vanish. This discrepancy signals an inconsistency between Eq. (9) and Eq. (11), highlighting the necessity of properly including the correction term \(\Delta_{n', \vec{k}'; n, \vec{k}}\), as derived in Eq. (13), where:
\begin{equation}
\vec{\Delta}_{n', \vec{k}'; n, \vec{k}}^H = \frac{\hbar^2}{2m} \int \nabla \left( \nabla f^*_{n', \vec{k}'} \nabla_{\vec{k}} f_{n, \vec{k}} - f^*_{n', \vec{k}'} \nabla \nabla_{\vec{k}} f_{n, \vec{k}} \right) d^3 r
\end{equation}

Alternatively, we may adopt the conventional framework in which \( H \) depends explicitly on \(\vec{k}\) via minimal substitution:
\begin{equation}
H(\vec{k}) = \frac{(\vec{p} + \hbar \vec{k})^2}{2m} + V(\vec{r}),
\end{equation}
with eigenfunctions \( u_{n, \vec{k}}(\vec{r}) \), the periodic cell functions. In this formulation, eigenfunctions corresponding to different values of \(\vec{k}\) are no longer eigenfunctions of the same operator. Therefore, there is no strict orthogonality between states \( u_{n', \vec{k}'} \) and \( u_{n, \vec{k}} \) for \( \vec{k}' \neq \vec{k} \), which is a critical point when evaluating matrix elements.

The apparent additional terms that arise in expressions involving derivatives with respect to \(\vec{k}\) are not due to a failure of Hermiticity, but instead stem from the domain issues associated with differentiating eigenfunctions of operators that change with the parameter. These domain subtleties are particularly evident when differentiating with respect to the crystal momentum \(\vec{k}\).

From the conventional formulation, we have:
\begin{equation}
\nabla_{\vec{k}} H(\vec{k}) = \frac{\hbar (\vec{p} + \hbar \vec{k})}{m}
\end{equation}

Consider now the general relation obtained from differentiating the eigenvalue equation (see Eq. (4)):
\begin{equation}
\langle u_{n', \vec{k}'} | \nabla_{\vec{k}} H(\vec{k}) | u_{n, \vec{k}} \rangle + \langle u_{n', \vec{k}'} | H(\vec{k}) \nabla_{\vec{k}} | u_{n, \vec{k}} \rangle = \epsilon_{n, \vec{k}} \langle u_{n', \vec{k}'} | \nabla_{\vec{k}} | u_{n, \vec{k}} \rangle
\end{equation}

However, since \( u_{n', \vec{k}'} \) is not an eigenfunction of \( H(\vec{k}) \), the second term must be modified to account for the fact that the operator \( H(\vec{k}) \) is acting outside its parameter-defined domain. We address this by writing:
\begin{equation}
\langle u_{n', \vec{k}'} | [H(\vec{k}) + H(\vec{k}') - H(\vec{k}')] \nabla_{\vec{k}} | u_{n, \vec{k}} \rangle
\end{equation}
which leads to:
\begin{align*}
\langle u_{n', \vec{k}'} | \nabla_{\vec{k}} H(\vec{k}) | u_{n, \vec{k}} \rangle 
+ \langle u_{n', \vec{k}'} | [H(\vec{k}) - H(\vec{k}')] \nabla_{\vec{k}} | u_{n, \vec{k}} \rangle \\
+ \langle u_{n', \vec{k}'} | H(\vec{k}') \nabla_{\vec{k}} | u_{n, \vec{k}} \rangle 
= \epsilon_{n, \vec{k}} \langle u_{n', \vec{k}'} | \nabla_{\vec{k}} | u_{n, \vec{k}} \rangle
\end{align*}

The contribution from \( H(\vec{k}') \nabla_{\vec{k}} \) must now be evaluated carefully. Since \( H(\vec{k}') \) acts on a function not in its domain, we account for the discrepancy via the correction term:
\begin{align}
\langle u_{n', \vec{k}'} | H(\vec{k}') \nabla_{\vec{k}} | u_{n, \vec{k}} \rangle 
= \langle H(\vec{k}') u_{n', \vec{k}'} | \nabla_{\vec{k}} | u_{n, \vec{k}} \rangle 
+ \Delta^{H(\vec{k}')}_{n', \vec{k}'; n, \vec{k}} \nonumber \\
= \epsilon_{n', \vec{k}'} \langle u_{n', \vec{k}'} | \nabla_{\vec{k}} | u_{n, \vec{k}} \rangle 
+ \Delta^{H(\vec{k}')}_{n', \vec{k}'; n, \vec{k}}
\end{align}

Thus, the corrected expression for the matrix element becomes:
\begin{align}
\langle u_{n', \vec{k}'} | \nabla_{\vec{k}} | u_{n, \vec{k}} \rangle = 
\frac{ \langle u_{n', \vec{k}'} | \nabla_{\vec{k}} H(\vec{k}) | u_{n, \vec{k}} \rangle }
{\epsilon_{n, \vec{k}} - \epsilon_{n', \vec{k}'} } 
+ \frac{ \langle u_{n', \vec{k}'} | [H(\vec{k}) - H(\vec{k}')] \nabla_{\vec{k}} | u_{n, \vec{k}} \rangle 
+ \Delta^{H(\vec{k}')}_{n', \vec{k}'; n, \vec{k}} }{\epsilon_{n, \vec{k}} - \epsilon_{n', \vec{k}'}}
\end{align}

Specializing to the case \(\vec{k}' = \vec{k}\), and evaluating \(\Delta_{n', \vec{k}; n, \vec{k}}^{H(\vec{k})}\) via Eq. (13), we obtain:
\begin{equation}
\vec{\Delta}_{n', n}^{H(\vec{k})} = \frac{\hbar^2}{2m} \int \nabla \left( 
\nabla u^*_{n', \vec{k}} \nabla_{\vec{k}} u_{n, \vec{k}} 
- u^*_{n', \vec{k}} \nabla \nabla_{\vec{k}} u_{n, \vec{k}} 
- 2i u^*_{n', \vec{k}} \vec{k} \nabla_{\vec{k}} u_{n, \vec{k}} 
\right) d^3 r
\end{equation}

After performing the necessary computations (see Appendix C), we arrive at the relation:
\begin{equation}
\vec{\Delta}_{n', n}^{H} - \vec{\Delta}_{n', \vec{k}; n, \vec{k}}^{H(\vec{k})} = 
\frac{\hbar}{m} \langle \vec{p} \rangle_{n', \vec{k}; n, \vec{k}} 
+ i (\epsilon_{n, \vec{k}} - \epsilon_{n', \vec{k}}) \langle \vec{r} \rangle_{n', \vec{k}; n, \vec{k}}
\end{equation}

This confirms the earlier result obtained in the 1D case (Eq. (23)). In that simpler case, \(\vec{\Delta}_{n', n}^{H(\vec{k})} = 0\) due to periodicity and boundary conditions on the \( u_{n, \vec{k}} \) functions. We expect this behavior to persist in higher dimensions (see discussion at the end of Appendix C), yielding:
\begin{equation}
\vec{\Delta}_{n', n}^{H} = 
\frac{\hbar}{m} \langle \vec{p} \rangle_{n', \vec{k}; n, \vec{k}} 
+ i (\epsilon_{n, \vec{k}} - \epsilon_{n', \vec{k}}) \langle \vec{r} \rangle_{n', \vec{k}; n, \vec{k}}
\end{equation}

Therefore, the Berry curvature (as given by Eq. (19)) for band \( n \) becomes:
\begin{align}
\vec{\Omega}^n(\vec{k}) = i \sum_{n' \ne n} \frac{ 
\vec{\Delta}_{n, n'}^{H} \times \vec{\Delta}_{n', n}^{H} 
}{(\epsilon_n(\vec{k}) - \epsilon_{n'}(\vec{k}))^2} \\
= i \sum_{n' \ne n} \frac{ 
\left( \frac{\hbar}{m} \langle \vec{p} \rangle_{n, n'} 
+ i (\epsilon_{n'} - \epsilon_n) \langle \vec{r} \rangle_{n, n'} \right) 
\times 
\left( \frac{\hbar}{m} \langle \vec{p} \rangle_{n', n} 
+ i (\epsilon_n - \epsilon_{n'}) \langle \vec{r} \rangle_{n', n} \right) 
}{(\epsilon_n - \epsilon_{n'})^2}
\end{align}

\section{Conclusions}

In this article, we have rigorously demonstrated the crucial role of accounting for domain-related subtleties in the calculation of Berry curvature, especially in scenarios where the Hamiltonian \( H \) appears to be independent of external parameters. This insight addresses several significant issues and reveals important implications for both theoretical and practical aspects of solid-state physics.

Our analysis shows that correctly capturing the behavior of Berry curvature requires careful attention to how the parameter dependence enters not only the Hamiltonian but also the eigenfunctions. When the Hamiltonian does not explicitly depend on parameters such as the crystal momentum \(\vec{k}\), conventional expressions for Berry curvature can yield incomplete or even misleading results. This is because the eigenstates themselves implicitly carry parameter dependence, and their variation must be treated carefully when evaluating derivatives such as \(\nabla_{\vec{k}}\). 

This previously overlooked aspect reveals that the correct treatment involves additional terms arising from domain subtleties — particularly in off-diagonal matrix elements involving derivatives with respect to parameters. These corrections restore consistency between different formulations and yield the physically correct Berry curvature, even when the Hamiltonian appears parameter-independent.

We have addressed the apparent paradox that arises in Bloch solid models, where the naive application of Berry curvature formulas might suggest a vanishing result due to the lack of explicit \(\vec{k}\)-dependence in the Hamiltonian. By examining how derivative operators act on parameter-dependent eigenstates and how domain considerations modify matrix elements, we showed that additional terms — such as \(\vec{\Delta}_{n', \vec{k}'; n, \vec{k}}^H\) — must be included to restore accuracy. These terms capture important contributions from position and momentum matrix elements and reconcile discrepancies in traditional derivations.

Our detailed mathematical treatment provides explicit expressions for these correction terms and demonstrates how they recover known physical results. Importantly, the analysis is not confined to one-dimensional models; the correction terms become even more significant in higher-dimensional systems where the parameter space is richer and the associated geometry more complex.

Ultimately, this study emphasizes the necessity of carefully treating the parameter dependence of quantum operators and their eigenstates, particularly when computing geometric quantities such as the Berry curvature. By doing so, we not only resolve inconsistencies in conventional treatments but also deepen our understanding of the geometric and topological aspects of quantum systems in solids.

Future research should continue to explore the implications of these domain-related effects, especially in systems with complex band structures or nontrivial topology. Accurate calculation of Berry curvature is essential for understanding a wide range of physical phenomena, from the anomalous Hall effect to the characterization of topological materials. Incorporating these insights into theoretical and computational frameworks will enable more precise predictions and a clearer understanding of emergent behaviors in modern condensed matter systems.

\section{Appendices}
\appendix

\section{Derivation of Equation (21)}
To derive Equation (21), we start by applying Equation (13) to a one-dimensional (1D) Bloch solid. Consider the Hamiltonian for this system:
\begin{equation}
H = \frac{p^2}{2m} + V(x),
\end{equation}
where \( V(x) \) is the periodic potential of the crystal. The eigenfunctions of this Hamiltonian are given by:
\begin{equation}
f_{n,k}(x) = u_{n,k}(x) \exp(i k x),
\end{equation}
where \( u_{n,k}(x) \) represents the periodic part of the Bloch function. To find the term \( \Delta_{k',k} \) in 1D, we apply the divergence theorem to Equation (13), leading to:
\begin{align}
\Delta_{k',k} &= \frac{\hbar^2}{2m} \left( \frac{\partial f_{n,k'}^*}{\partial x} \frac{\partial f_{n,k}}{\partial k} - f_{n,k'}^* \frac{\partial}{\partial x} \frac{\partial f_{n,k}}{\partial k} \right) \\
&= \frac{\hbar^2}{2m} \left[ (k+k') x u_{k'}^* u_k + i x \left( u_k \frac{\partial u_{k'}^*}{\partial x} - u_{k'}^* \frac{\partial u_k}{\partial x} \right) \bigg|_{0}^{L} \right].
\end{align}
This result can be related to the off-diagonal matrix elements of the momentum and position operators. Specifically, we have:
\begin{align}
\langle p \rangle_{k'k} &= -i \hbar \int_{0}^{L} dx \, e^{i (k - k') x} u_{k'}^* \frac{\partial u_k}{\partial x} \\
&= \frac{i \hbar}{2} x e^{i (k - k') x} \left( u_k \frac{\partial u_{k'}^*}{\partial x} - u_{k'}^* \frac{\partial u_k}{\partial x} \right) \bigg|_{0}^{L} \\
& \quad + \frac{i \hbar}{2} \int_{0}^{L} dx \, e^{i (k - k') x} \left[ i (k - k') \left( u_{k'}^* \frac{\partial u_k}{\partial x} - u_k \frac{\partial u_{k'}^*}{\partial x} \right) \right. \\
& \left. + u_{k'}^* \frac{\partial^2 u_k}{\partial x^2} - u_k \frac{\partial^2 u_{k'}^*}{\partial x^2} \right].
\end{align}
Using the Schrödinger equation, we can eliminate the second derivatives as follows:
\begin{align}
\frac{\partial^2 u_k}{\partial x^2} &= -\frac{2m}{\hbar^2} (\epsilon_k - V(x)) u_k + k^2 u_k - 2ik \frac{\partial u_k}{\partial x}, \\
\frac{\partial^2 u_{k'}^*}{\partial x^2} &= -\frac{2m}{\hbar^2} (\epsilon_{k'} - V(x)) u_{k'}^* + k'^2 u_{k'}^* + 2ik' \frac{\partial u_{k'}^*}{\partial x}.
\end{align}
Substituting these expressions into the integral results in:
\begin{align}
\langle p \rangle_{k'k} &= \frac{im}{\hbar} (\epsilon_{k'} - \epsilon_k) \langle x \rangle_{k'k} \\
& \quad + \frac{\hbar}{2} \left[ (k + k') x e^{i (k - k') x} u_{k'}^* u_k \right. \\
& \left. + i x e^{i (k - k') x} \left( u_k \frac{\partial u_{k'}^*}{\partial x} - u_{k'}^* \frac{\partial u_k}{\partial x} \right) \bigg|_{0}^{L} \right].
\end{align}
By comparing this result with Equation (54), we find:
\begin{equation}
\Delta_{k',k} = \frac{\hbar}{m} \left[ \langle p \rangle_{k',k} + \frac{im}{\hbar} (\epsilon_k - \epsilon_{k'}) \langle x \rangle_{k',k} \right].
\end{equation}

\section{Proof that Equation (17) Has Zero Divergence}
We start by considering the following operator:
\begin{equation}
\vec{B} = -i \sum_k | \nabla_{\vec{R}} k \rangle \langle k |.
\end{equation}
To show that \( \vec{B} \) is a Hermitian operator, we examine the matrix elements:
\begin{align}
\langle m | \vec{B} | n \rangle &= -i \langle m | \nabla_{\vec{R}} n \rangle, \\
\langle \vec{B} m | n \rangle &= -i \langle m | \nabla_{\vec{R}} n \rangle \\
&= i \langle \nabla_{\vec{R}} m | n \rangle \\
&= -i \langle m | \nabla_{\vec{R}} n \rangle.
\end{align}
Thus, the matrix elements are:
\begin{equation}
\langle m | \vec{B} | n \rangle = -i \langle m | \nabla_{\vec{R}} n \rangle = -i \frac{\langle m | \nabla_{\vec{R}} H | n \rangle + \vec{\Delta}_{m,n}}{\epsilon_n(\vec{R}) - \epsilon_m(\vec{R})}.
\end{equation}
Similarly, we have:
\begin{equation}
\langle n | \vec{B} | m \rangle = -i \langle n | \nabla_{\vec{R}} m \rangle = -i \frac{\langle n | \nabla_{\vec{R}} H | m \rangle + \vec{\Delta}_{n,m}}{\epsilon_m(\vec{R}) - \epsilon_n(\vec{R})}.
\end{equation}
Thus, Equation (10) becomes:
\begin{align}
\vec{\Omega} &= i \sum_{m \ne n} \langle \nabla_{\vec{R}} n | m \rangle \times \langle m | \nabla_{\vec{R}} n \rangle \\
&= -i \sum_{m \ne n} \langle n | \nabla_{\vec{R}} m \rangle \times \langle m | \nabla_{\vec{R}} n \rangle \\
&= i \langle n | \vec{B} \times \vec{B} | n \rangle.
\end{align}
To find the divergence of \( \vec{\Omega} \), we use:
\begin{align}
\nabla_{\vec{R}} \cdot \vec{\Omega} &= i \langle \nabla_{\vec{R}} n | \vec{B} \times \vec{B} | n \rangle + i \langle n | \vec{B} \times \vec{B} | \nabla_{\vec{R}} n \rangle \\
& \quad + i \langle n | \nabla_{\vec{R}} \cdot (\vec{B} \times \vec{B}) | n \rangle.
\end{align}
From Equation (67), we have:
\begin{equation}
| \nabla_{\vec{R}} n \rangle = i \vec{B} | n \rangle.
\end{equation}
Thus:
\begin{align}
\nabla_{\vec{R}} \cdot \vec{\Omega} &= \langle n | \vec{B} \cdot (\vec{B} \times \vec{B}) | n \rangle - \langle n | (\vec{B} \times \vec{B}) \cdot \vec{B} | n \rangle \\
& \quad + i \langle n | \nabla_{\vec{R}} \cdot (\vec{B} \times \vec{B}) | n \rangle.
\end{align}
For the divergence of \( \vec{B} \times \vec{B} \), we have:
\begin{align}
\nabla_{\vec{R}} \cdot (\vec{B} \times \vec{B}) &= (\nabla_{\vec{R}} \times \vec{B}) \cdot \vec{B} + \vec{B} \cdot (\nabla_{\vec{R}} \times \vec{B}).
\end{align}
From Equation (67):
\begin{equation}
\nabla_{\vec{R}} \times \vec{B} = i \vec{B} \times \vec{B}.
\end{equation}
Therefore:
\begin{equation}
\nabla_{\vec{R}} \cdot \vec{\Omega} = 0,
\end{equation}
except at points of degeneracy.

\section{Proof of Eq. 47: The difference of Delta's in the two regimes}
Consider the Equation (39) (for $\vec{k'}=\vec{k}$):
\begin{equation}
\vec{\Delta}^{H}_{n', n}(\vec{k}) = \frac{\hbar^2}{2m} \int d^3r \, \nabla \left( \nabla f_{n'}^* \nabla_{\vec{k}} f_n - f_{n'}^* \nabla \nabla_{\vec{k}} f_n \right)
\end{equation}
Expanding the gradient terms:
\begin{equation}
\nabla f_{n'}^* = -i \vec{k} e^{-i \vec{k} \cdot \vec{r}} u_{n', \vec{k}}^* + e^{-i \vec{k} \cdot \vec{r}} \nabla u_{n', \vec{k}}^*
\end{equation}
\begin{equation}
\nabla_{\vec{k}} f_n = i \vec{r} e^{i \vec{k} \cdot \vec{r}} u_{n, \vec{k}} + e^{i \vec{k} \cdot \vec{r}} \nabla_{\vec{k}} u_{n, \vec{k}}
\end{equation}
The second gradient $\nabla \nabla_{\vec{k}} f_n$ expands to the following:
\begin{align}
\nabla \nabla_{\vec{k}} f_n = & \, i \hat{i} e^{i \vec{k} \cdot \vec{r}} u_{n, \vec{k}} + i \hat{j} e^{i \vec{k} \cdot \vec{r}} u_{n, \vec{k}} + i \hat{k} e^{i \vec{k} \cdot \vec{r}} u_{n, \vec{k}} \nonumber \\
& - \vec{r} \vec{k} e^{i \vec{k} \cdot \vec{r}} u_{n, \vec{k}} \nonumber \\
& + i \vec{r} e^{i \vec{k} \cdot \vec{r}} \nabla u_{n, \vec{k}} + i \vec{k} e^{i \vec{k} \cdot \vec{r}} \nabla_{\vec{k}} u_{n, \vec{k}} \nonumber \\
& + e^{i \vec{k} \cdot \vec{r}} \nabla \nabla_{\vec{k}} u_{n, \vec{k}}
\end{align}
Thus, we obtain:
\begin{align}
\vec{\Delta}^{H}_{n', n}(\vec{k})  = \frac{\hbar^2}{2m} \int d^3r \, \nabla \Big[ & 2 \vec{k} u_{n', \vec{k}}^* \vec{r} u_{n, \vec{k}} - 2i \vec{k} u_{n', \vec{k}}^* \nabla_{\vec{k}} u_{n, \vec{k}} + i \vec{r} u_{n, \vec{k}} \nabla u_{n', \vec{k}}^* \nonumber \\
& - i \vec{r} u_{n', \vec{k}}^* \nabla u_{n, \vec{k}} + \nabla_{\vec{k}} u_{n, \vec{k}} \nabla u_{n', \vec{k}}^* - u_{n', \vec{k}}^* \nabla \nabla_{\vec{k}} u_{n, \vec{k}} \nonumber \\
& - i u_{n, \vec{k}} u_{n', \vec{k}}^* (\hat{i} \hat{i} + \hat{j} \hat{j} + \hat{k} \hat{k}) \Big]
\end{align}
We can divide this into two parts:
\begin{align}
\vec{\Delta}^{H}_{n', n}(\vec{k}) = \frac{\hbar^2}{2m} \int d^3r \, \nabla \Big[ & -2i \vec{k} u_{n', \vec{k}}^* \nabla_{\vec{k}} u_{n, \vec{k}} + \nabla_{\vec{k}} u_{n, \vec{k}} \nabla u_{n', \vec{k}}^* - u_{n', \vec{k}}^* \nabla \nabla_{\vec{k}} u_{n, \vec{k}} \Big] \nonumber \\
+ \frac{\hbar^2}{2m} \int d^3r \, \nabla \Big[ & 2 \vec{k} u_{n', \vec{k}}^* \vec{r} u_{n, \vec{k}} + i \vec{r} u_{n, \vec{k}} \nabla u_{n', \vec{k}}^* \nonumber \\
& - i \vec{r} u_{n', \vec{k}}^* \nabla u_{n, \vec{k}} - i u_{n, \vec{k}} u_{n', \vec{k}}^* (\hat{i} \hat{i} + \hat{j} \hat{j} + \hat{k} \hat{k}) \Big]
\end{align}
Let us now focus on the second additive term of Eq. (88).  The first term gives:
\begin{align}
    & \frac{\hbar^2}{2m} \int d^3 r \nabla \left[ 2\vec{k} u_{n', \vec{k}}^* \vec{r} u_{n, \vec{k}} \right] \nonumber \\
    & = \frac{\hbar^2}{m} \int d^3 r \left( k_x \frac{\partial}{\partial x} \left[ u_{n', \vec{k}}^* \vec{r} u_{n, \vec{k}} \right] + k_y \frac{\partial}{\partial y} \left[ u_{n', \vec{k}}^* \vec{r} u_{n, \vec{k}} \right] + k_z \frac{\partial}{\partial z} \left[ u_{n', \vec{k}}^* \vec{r} u_{n, \vec{k}} \right] \right) \nonumber \\
    & = \frac{\hbar^2}{m} \int d^3 r \vec{r} \left[ \vec{k} \nabla u_{n', \vec{k}}^* u_{n, \vec{k}} + u_{n', \vec{k}}^* \vec{k} \nabla u_{n, \vec{k}} \right]. 
\end{align}
The second term gives:
\begin{align}
    & \int d^3 r \nabla \left[ i \vec{r} u_{n, \vec{k}} \nabla u_{n', \vec{k}}^* - i \vec{r} u_{n', \vec{k}}^* \nabla u_{n, \vec{k}} \right] \nonumber \\
    & = \int d^3 r \left[ i \hat{i} u_{n, \vec{k}} \nabla u_{n', \vec{k}}^* + i \hat{j} u_{n, \vec{k}} \nabla u_{n', \vec{k}}^* + i \hat{k} u_{n, \vec{k}} \nabla u_{n', \vec{k}}^* \right. \nonumber \\
    & \quad \left. + i \vec{r} \nabla u_{n, \vec{k}} \nabla u_{n', \vec{k}}^* + i \vec{r} u_{n, \vec{k}} \nabla^2 u_{n', \vec{k}}^* \right. \nonumber \\
    & \quad \left. - i \hat{i} u_{n', \vec{k}}^* \nabla u_{n, \vec{k}} - i \hat{j} u_{n', \vec{k}}^* \nabla u_{n, \vec{k}} - i \hat{k} u_{n', \vec{k}}^* \nabla u_{n, \vec{k}} \right. \nonumber \\
    & \quad \left. - i \vec{r} \nabla u_{n', \vec{k}}^* \nabla u_{n, \vec{k}} - i \vec{r} u_{n', \vec{k}}^* \nabla^2 u_{n, \vec{k}} \right].
\end{align}
The third term gives:
\begin{align}
    & \int d^3 r \left[-i \nabla u_{n, \vec{k}} u_{n', \vec{k}}^* (\hat{i} + \hat{j} + \hat{k}) - i u_{n, \vec{k}} \nabla u_{n', \vec{k}}^* (\hat{i} + \hat{j} + \hat{k}) \right].
\end{align}
Combining the terms we get:
\begin{align}
    & \frac{\hbar^2}{2m} \int d^3 r \nabla \left[ 2 \vec{k} u_{n', \vec{k}}^* \vec{r} u_{n, \vec{k}} + i \vec{r} u_{n, \vec{k}} \nabla u_{n', \vec{k}}^* - i \vec{r} u_{n', \vec{k}}^* \nabla u_{n, \vec{k}} \right. \nonumber \\
    & \quad \left. - i u_{n, \vec{k}} u_{n', \vec{k}}^* (\hat{i} + \hat{j} + \hat{k}) \right] \nonumber \\
    & = \frac{\hbar^2}{m} \int d^3 r \vec{r} \left[ \vec{k} \nabla u_{n', \vec{k}}^* u_{n, \vec{k}} + u_{n', \vec{k}}^* \vec{k} \nabla u_{n, \vec{k}} \right] \nonumber \\
    & - \frac{i \hbar^2}{m} \int d^3 r \left[ u_{n', \vec{k}}^* \nabla u_{n, \vec{k}} \right] \nonumber \\
    & + \frac{i \hbar^2}{2m} \int d^3 r \vec{r} \left[ u_{n, \vec{k}} \nabla^2 u_{n', \vec{k}}^* - u_{n', \vec{k}}^* \nabla^2 u_{n, \vec{k}} \right].
\end{align}
Consider now the term 
\begin{equation}
    u_{n, \vec{k}} \nabla^2 u_{n', \vec{k}}^* - u_{n', \vec{k}}^* \nabla^2 u_{n, \vec{k}}.
\end{equation}
From the Schrödinger equation, we get:
\begin{align}\nabla^2 u_{n', \vec{k}}^* &= k^2 u_{n', \vec{k}}^* + 2i\vec{k} \cdot \nabla u_{n', \vec{k}}^* - \frac{2m}{\hbar^2} \varepsilon_{n'} u_{n', \vec{k}}^* + \frac{2m}{\hbar^2} V(\vec{r}) u_{n', \vec{k}}^*, \\
    \nabla^2 u_{n, \vec{k}} &= k^2 u_{n, \vec{k}} - 2i\vec{k} \cdot \nabla u_{n, \vec{k}} - \frac{2m}{\hbar^2} \varepsilon_n u_{n, \vec{k}} + \frac{2m}{\hbar^2} V(\vec{r}) u_{n, \vec{k}}\end{align}
Thus, we have:
\begin{align}
    u_{(n, \vec{k})} \nabla^2 u_{n', \vec{k}}^* - u_{n', \vec{k}}^* \nabla^2 u_{n, \vec{k}} &= 2i u_{n, \vec{k}} \vec{k} \cdot \nabla u_{n', \vec{k}}^* + 2i u_{n', \vec{k}}^* \vec{k} \cdot \nabla u_{n, \vec{k}} \nonumber \\
    & \quad + \frac{2m}{\hbar^2} (\varepsilon_n - \varepsilon_{n'}) u_{n', \vec{k}}^* u_{n, \vec{k}}
\end{align}
Now, putting everything together, we conclude that the term
\begin{equation}
    \frac{\hbar^2}{2m} \int d^3 r \nabla \left[ 2\vec{k} u_{(n', \vec{k})}^* \vec{r} u_{n, \vec{k}} + i\vec{r} u_{n, \vec{k}} \nabla u_{n', \vec{k}}^* - i\vec{r} u_{n', \vec{k}}^* \nabla u_{n, \vec{k}} - i u_{n, \vec{k}} u_{n', \vec{k}}^* (\hat{i} + \hat{j} + \hat{k}) \right]
\end{equation}
is
\begin{align}
    &= -\frac{i\hbar^2}{m} \int d^3 r \left[ u_{n', \vec{k}}^* \nabla u_{n, \vec{k}} \right] + \frac{\hbar^2}{m} \frac{im}{\hbar^2} (\varepsilon_n - \varepsilon_{n'}) \int d^3 r \left[ \vec{r} u_{n', \vec{k}}^* u_{n, \vec{k}} \right]
\end{align}
and 
\begin{equation}
\vec{\Delta}^{H}_{n', n}(\vec{k}) = -\frac{i\hbar^2}{m} \int d^3 r \left[ u_{n', \vec{k}}^* \nabla u_{n, \vec{k}} \right] + \frac{\hbar^2}{m} \frac{im}{\hbar^2} (\varepsilon_n - \varepsilon_{n'}) \int d^3 r \left[ \vec{r} u_{n', \vec{k}}^* u_{n, \vec{k}} \right] 
\end{equation}
\begin{align}
    &\quad + \frac{\hbar^2}{2m} \int d^3 r \nabla \left[ -2i\vec{k} u_{n', \vec{k}}^* \nabla_{\vec{k}} u_{n, \vec{k}} + \nabla_{\vec{k}} u_{n, \vec{k}} \nabla u_{n', \vec{k}}^* - u_{n', \vec{k}}^* \nabla \nabla_{\vec{k}} u_{n, \vec{k}} \right]
\end{align}

On the other hand, for the `conventional' Hamiltonian Bloch Solid we have:
\begin{equation}
    H = \frac{(\vec{p} + \hbar \vec{k})^2}{2m} + V(\vec{r}),
\end{equation}
and wavefunctions/vector potential $\vec{A}$:
\begin{equation}
    f_{(n, \vec{k})} = u_{(n, \vec{k})}, \quad \vec{A} = \frac{\hbar c}{e} \vec{k},
\end{equation}
Equation (17) then becomes:
\begin{equation}
    \Omega_n(\vec{k}) = i \sum_{n' \neq n} \frac{(\langle n | \nabla_{\vec{k}} H | n' \rangle + \vec{\Delta}^{H(\vec{k})}_{n, n'}) \times (\langle n' | \nabla_{\vec{k}} H | n \rangle +  \vec{\Delta}^{H(\vec{k})}_{n', n})}{(\varepsilon_n - \varepsilon_{n'})^2}.
\end{equation}
If we now examine the term 
\begin{equation}
    \langle n' | \nabla_{\vec{k}} H | n \rangle + \vec{\Delta}^{H(\vec{k})}_{n', n}
\end{equation}
with 
\begin{equation}
    \vec{\Delta}^{H(\vec{k})}_{n', n} = \frac{\hbar^2}{2m} \int d^3 r \nabla \left( \nabla f_{n'}^* \nabla_{\vec{k}} f_n - f_{n'}^* \nabla \nabla_{\vec{k}} f_n - \frac{2i e}{\hbar c} f_{n'}^* \vec{A} \nabla_{\vec{k}} f_n \right),
\end{equation}
and 
\begin{align}
    \langle n' | \nabla_{\vec{k}} H | n \rangle &= \int d^3 r \, u_{n', \vec{k}}^* \frac{\hbar(\vec{p} + \hbar \vec{k})}{m} u_{n, \vec{k}} \nonumber \\
    &= \frac{\hbar}{m} \int d^3 r \, u_{n', \vec{k}}^* \left(-i\hbar \nabla + \hbar \vec{k}\right) u_{n, \vec{k}} \nonumber \\
    &= -i \frac{\hbar^2}{m} \int d^3 r \, u_{n', \vec{k}}^* \nabla u_{n, \vec{k}}.
\end{align}
Thus, we have 
\begin{align}
  \vec{\Delta}^{H(\vec{k})}_{n', n} &= \frac{\hbar^2}{2m} \int d^3 r \nabla \left( \nabla u_{n', \vec{k}}^* \nabla_{\vec{k}} u_{n, \vec{k}} - u_{n', \vec{k}}^* \nabla \nabla_{\vec{k}} u_{n, \vec{k}} - 2i \vec{k} u_{n', \vec{k}}^* \nabla_{\vec{k}} u_{n, \vec{k}} \right).
\end{align}
Now, we combine these results to find:
\begin{align}
    \langle n' | \nabla_{\vec{k}} H | n \rangle +  \vec{\Delta}^{H(\vec{k})}_{n', n} &= -i \frac{\hbar^2}{m} \int d^3 r \, u_{n', \vec{k}}^* \nabla u_{n, \vec{k}} \nonumber \\
    & \quad + \frac{\hbar^2}{2m} \int d^3 r \nabla \bigg[ 
        \nabla u_{n', \vec{k}}^* \nabla_{\vec{k}} u_{n, \vec{k}} \nonumber \\
    & \quad \quad - u_{n', \vec{k}}^* \nabla \nabla_{\vec{k}} u_{n, \vec{k}} 
    - 2i \vec{k} u_{n', \vec{k}}^* \nabla_{\vec{k}} u_{n, \vec{k}} 
    \bigg].
\end{align}
Now, if we compare this term 
\begin{equation}
    \langle n' | \nabla_{\vec{k}} H | n \rangle +  \vec{\Delta}^{H(\vec{k})}_{n', n}
\end{equation}
with Eq. (99) previously calculated, one can clearly see that their difference $  \langle n' | \nabla_{\vec{k}} H | n \rangle +  \vec{\Delta}^{H(\vec{k})}_{n', n} - \vec{\Delta}^{H}_{n', n}$ is just the term 
\begin{equation}
   - i(\varepsilon_n - \varepsilon_{(n')}) \int d^3 r \, [\vec{r} u_{(n', \vec{k})}^* u_{n, \vec{k}}] \quad (\text{mean value of the position operator}),
\end{equation}
finally proving Eq. (47):
\begin{equation} \label{eq:Delta_difference}
\begin{aligned}
    \vec{\Delta}^{H}_{n', n} - \vec{\Delta}^{H(\vec{k})}_{n', n} &= \langle n' | \nabla_{\vec{k}} H | n \rangle \\
    &\quad + i(\varepsilon_n - \varepsilon_{n'}) \int d^3 r \, [\vec{r} u_{n', \vec{k}}^* u_{n, \vec{k}}] \\
    &= \left[ \frac{\hbar}{m} \langle \vec{p} \rangle_{n',  n} + i (\epsilon_{n, \vec{k}} - \epsilon_{n', \vec{k}}) \langle \vec{r} \rangle_{n', n} \right]
\end{aligned}
\end{equation}
The term $\vec{\Delta}^{H(\vec{k})}_{n', n}$ is anticipated to be zero because it involves the derivatives of periodic cell functions, which exhibit a fundamental property of periodicity. In a periodic potential, such as those found in crystal structures, the wavefunctions are typically periodic in nature, leading to the conclusion that their derivatives over a complete periodic cell will average out to zero. To illustrate this concept more clearly, we can examine a one-dimensional analog of this situation. In one dimension, if we consider a periodic function, its integral over one complete period yields a net result of zero when we compute the derivative with respect to the wavevector $\vec{k}$. Thus, just as the one-dimensional case demonstrates that the contributions of periodic functions vanish upon differentiation, we can extend this reasoning to higher dimensions, confirming that $\vec{\Delta}^{H(\vec{k})}_{n', n}$ is indeed expected to be zero. In the one-dimensional case, the expression for \(\Delta_{n',n}^{H(k)}\) is given by 
\[
\Delta_{n',n}^{H(k)} = \frac{\hbar^2}{2m} \left( \frac{\partial u_{(n', k)}^*}{\partial x} \frac {\partial u_{(n, k)}}{\partial k} - u_{(n', k)}^* \frac {\partial}{\partial x}\frac{\partial u_{(n, k)}}{\partial k}- 2i k u_{(n', k)}^* \frac{\partial u_{(n, k)}}{\partial k} \right) \bigg|_0^L.
\]
To evaluate this expression over the interval \([0, L]\), we observe that the terms involve derivatives of the wavefunctions that are periodic functions. Specifically, because the wavefunctions \(u_{(n,k)}\) and \(u_{(n',k)}\) are solutions to the Schrödinger equation in a periodic potential, they can be expressed as periodic functions of the form \(u(x) = u(x + L)\). As a result, the evaluation of the derivatives at the boundaries \(0\) and \(L\) will yield:
\begin{enumerate}
    \item \textbf{First Term:} The term $\frac{\partial u_{(n', k)}^*}{\partial x} \frac {\partial u_{(n, k)}}{\partial k}$ will evaluate to the same value at both boundaries \(0\) and \(L\), thus giving zero contribution.
    \item \textbf{Second Term:} The term $u_{(n', k)}^* \frac {\partial}{\partial x}\frac{\partial u_{(n, k)}}{\partial k}$ will similarly yield the same value at both boundaries, leading again to a zero contribution.
    \item \textbf{Third Term:} The term $2i k u_{(n', k)}^* \frac{\partial u_{(n, k)}}{\partial k} $ will also vanish for the same reason, as both wavefunctions are periodic.
\end{enumerate}
Since all terms evaluate to zero when considering their contributions from \(0\) to \(L\), we find that \(\Delta_{n',n}^{H(k)}\) is indeed zero in one dimension.
\textbf{Extension to Three Dimensions:} This reasoning can be extended to three dimensions based on similar periodicity arguments. In three dimensions, the wavefunctions $u_{n,\vec{k}}$ and $u_{n',\vec{k}}$ will still possess the periodic properties characteristic of solutions to the Schrödinger equation in periodic potentials. The gradients involved in  \(\vec{\Delta}_{n',n}^{H(\vec{k})}\) will similarly evaluate to zero over a complete unit cell in three-dimensional space due to the inherent periodicity. Therefore, we can confidently assert that the contributions from the periodic derivatives in three dimensions will also average out to zero, leading to the conclusion that \(\vec{\Delta}_{n',n}^{H(\vec{k})}\) is expected to remain zero in a three-dimensional scenario as well.

\end{document}